\newcommand{\algaas}{$\rm Al_{x}\-Ga_{1-x}\-As$}

\newcommand{\gaasalgaas}{$\rm GaAs/Al_{x}\-Ga_{1-x}\-As$}

\newcommand{\ingaasp}{$\rm InGa_{0.53x}\-As_{x}\-P$}

\newcommand{\figcap}[1]{\caption{\protect \footnotesize #1}}
\newcommand{\smalleps}[3]{\begin{figure}[#1]\centerline{ 
   \epsfxsize=7.9cm
   \epsfysize=7.6cm
   \epsffile{#2}} 
   \figcap{#3} 
   \end{figure}} 
\documentclass[twocolumn]{munich}
\usepackage{epsf,emlines,fleqn}
\begin{document}
\setlength{\paperwidth}{21cm} \setlength{\paperheight}{29.7cm}
\setlength{\topmargin}{0.4cm} \setlength{\oddsidemargin}{-0.2cm} 
\parindent 3ex

\pagestyle{empty}
\baselineskip 1em

\begin{frontmatter}

\title{\large \bf Modelling Multi Quantum Well Solar Cell Efficiency }
\author[London]{James P. Connolly$^{\rm 1}$}
\author[London]{Jenny Nelson}
\author[London]{Ian Ballard}
\author[London]{Keith W.J. Barnham}
\author[London]{Carsten Rohr}
\author[Sheffield]{Chris Button}
\author[Sheffield]{John Roberts}
\author[Nottingham]{Tom Foxon}

\vspace{-0.1cm}
\address[London]{Blackett Laboratory, Imperial College of
Science, Technology and Medicine, London SW7 2BZ }
\thanks{Electronic mail: j.connolly@ic.ac.uk}
\address[Sheffield]{ EPSRC III-V Facility, University of Sheffield,
Sheffield S1 3JD UK}
\address[Nottingham]{Physics Dept., University of Nottingham, Nottingham
N67 2RD UK}

\vspace{-0.16cm}

\begin{keyword}
\vspace{-2\baselineskip}
    Quantum wells - 1: Solar Cell Efficiencies - 2: Modelling - 3
\end{keyword}

\begin{abstract}

 ABSTRACT: The spectral response of quantum well solar cells (QWSCs) 
 is well understood.  We describe work on QWSC dark current theory 
 which combined with SR theory yields a system efficiency.  A 
 methodology published for single quantum well (SQW) systems is 
 extended to MQW systems in the \algaas\ and \ingaasp\ systems.  The 
 materials considered are dominated by Shockley-Read-Hall (SRH) 
 recombination.  The SRH formalism expresses the dark current in terms 
 of carrier recombination through mid-gap traps.  The SRH 
 recombination rate depends on the electron and hole densities of 
 states (DOS) in the barriers and wells, which are well known, and of 
 carrier non-radiative lifetimes.  These material quality dependent 
 lifetimes are extracted from analysis of suitable bulk control 
 samples.  Consistency over a range of AlGaAs controls and QWSCs is 
 examined, and the model is applied to QWSCs in InGaAsP on InP 
 substrates.  We find that the dark currents of MQW systems require a 
 reduction of the quasi Fermi level separation between carrier 
 populations in the wells relative to barrier material, in line with 
 previous studies.  Consequences for QWSCs are considered suggesting a 
 high efficiency potential.
\end{abstract}

\end{frontmatter}

\section{Introduction}
The QWSC \cite{barnham90} is a p-i-n structure with narrow regions or 
quantum wells  (QWs) of lower bandgap sandwiched between barrier layers with 
higher bandgap.  Both are situated in the i region and subject to a 
field in the operating regime.  The wells are narrow enough that 
carriers generated in them can only occupy discrete energy levels.

The control structures discussed here are p-i-n devices of similar 
material composition and dimensions as the QWSCs but with the QWs 
replaced with bulk material of the same composition as the barriers 
for barrier controls, or the barriers replaced with well material for 
well controls.  In some cases \ingaasp\ p-i-n heterostructures where 
the i region is made of material with either the confined well or the 
barrier bandgap are used as controls.

Experiment has shown (\cite{paxman93}, \cite{nelsonthinfilms}) that 
light absorbed in the QWs is converted to photocurrent with 
essentially unit efficiency, indicating good photogenerated carrier 
escape and collection efficiency.

Work on the efficiency of this system has raised a number of 
questions.  It has been suggested \cite{Araujo94} that the cell can be 
no more efficienct than a homostructure with an equivalent bandgap in 
the ideal limit and that the $\rm V_{oc}$ is determined by the lowest 
bandgap in the cell.  In real structures however further work
(\cite{voltageenhancement}, \cite{nelson96}, \cite{Tsui96}) has indicated
that assumptions  regarding a constant quasi-Fermi level separation
$\Delta E_{f}$ in  the i region may not be true, in which case higher
efficiencies are possible with this system.

This work investigates a QWSC dark current model dominated by 
Shockley-Read-Hall (SRH) recombination \cite{srh52} in a number of QWSCs and 
control cells without wells in two different material systems with a 
view to exploring the efficiency of this system.

\smalleps{htbp}{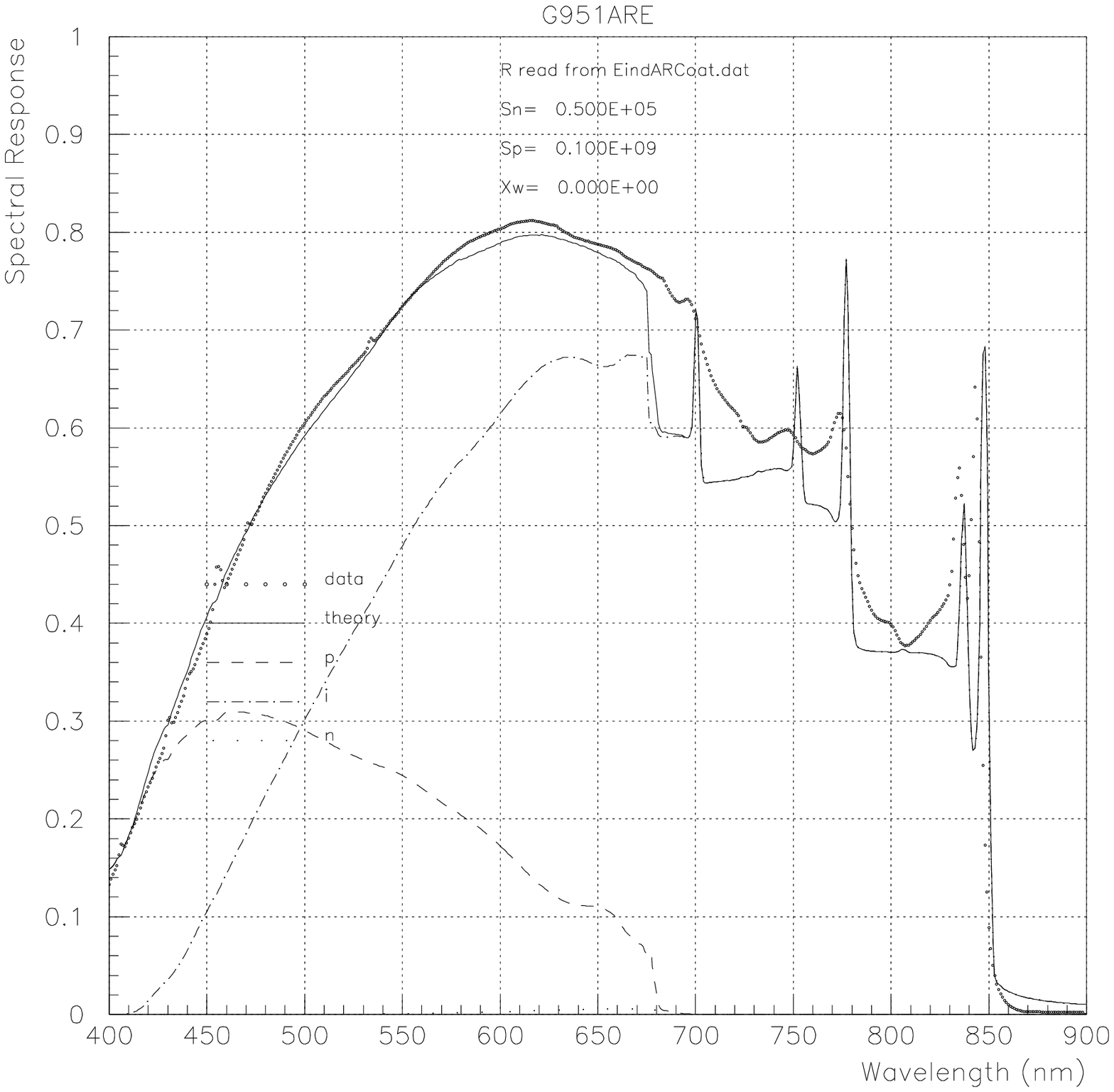}{Theory and experiment for a 50 
well $\rm Al_{36}Ga_{0.64}As$ QWSC with GaAs 
wells\label{qealgaasqwsc}}
\smalleps{htbp}{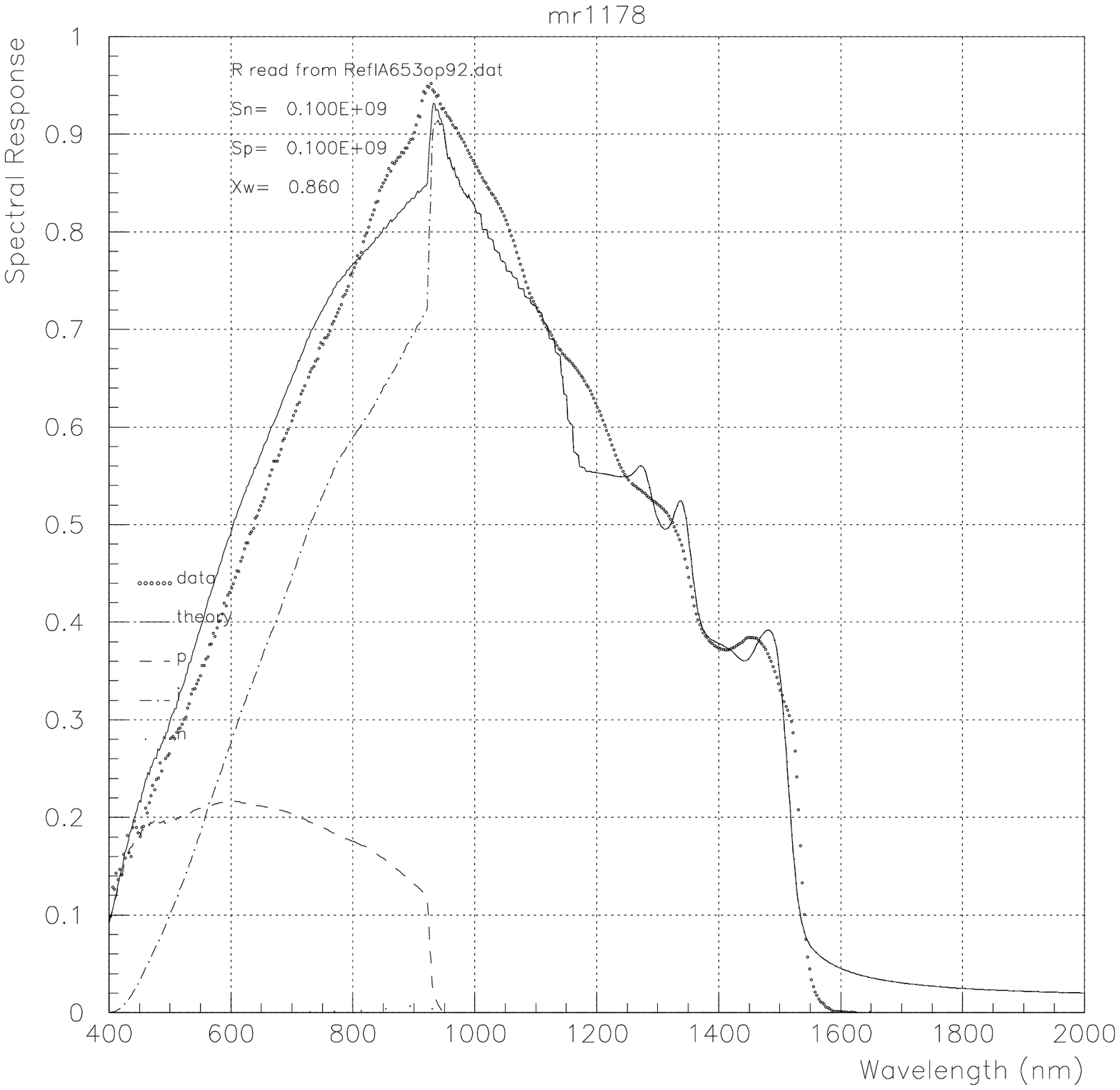}{Theory and experiment for a 30 
well $\rm InGaAsP$ QWSC with InGaAs wells\label{qeingaaspqwsc}}

\section{Model}

\subsection{Photocurrent}

The modelling of the spectral response (SR) has been discussed 
previously \cite{paxman93} but is summarised here.  It proceeds via 
solutions to the transport equations for minority carriers in the 
charge neutral regions of the cell.  Firstly these require information 
on minority carrier generation.  The bulk absorption coefficient is 
expressed via a non linear interpolation between published data sets 
and determines the generation rate.  The photocurrent from p and n 
regions is found by solving carrier current and continuity equations 
subject to minority carrier transport constants drawn from the 
literature.  The surface recombination velocity and minority carrier 
diffusion lengths are the device dependent fitting parameters.  The 
surface recombination velocity can be extracted from the short 
wavelength response, and the diffusion length from the broad shape of 
the spectral response for the p or n contribution as appropriate.

The intrinsic region calculation assumes 100\% collection for carriers
generated within it, which matches observation closely.  For 
contributions from the well, the above barrier bandgap contribution is 
given by the absorption of the quantum well material in the bulk.  
Below the barrier bandgap for the quantum well levels, the absorption 
is calculated from first principles assuming infinite barrier 
thickness and using published values of well and barrier bandgaps in 
the bulk, and electron and hole effective masses by solution of the 
Schr\"odinger equation in the effective mass and envelope function 
approximations, and excitonic states are included \cite{nelsonthinfilms}.

SR theory and data for a 50 well GaAs/AlGaAs QWSC and a 30 well 
InGaAsP on an InP substrate are shown in figures \ref{qealgaasqwsc} 
and \ref{qeingaaspqwsc} respectively.  The approximations hold 
satisfactorily though increasing underestimation is visible near the 
top of the well in the region of transition from 2D to a 3D density of 
states (DOS).  Overall the good fits show that the quantum well DOS is 
well described and that the assumption of unit escape efficiency for 
carrier escape from the wells is accurate.

\smalleps{htbp}{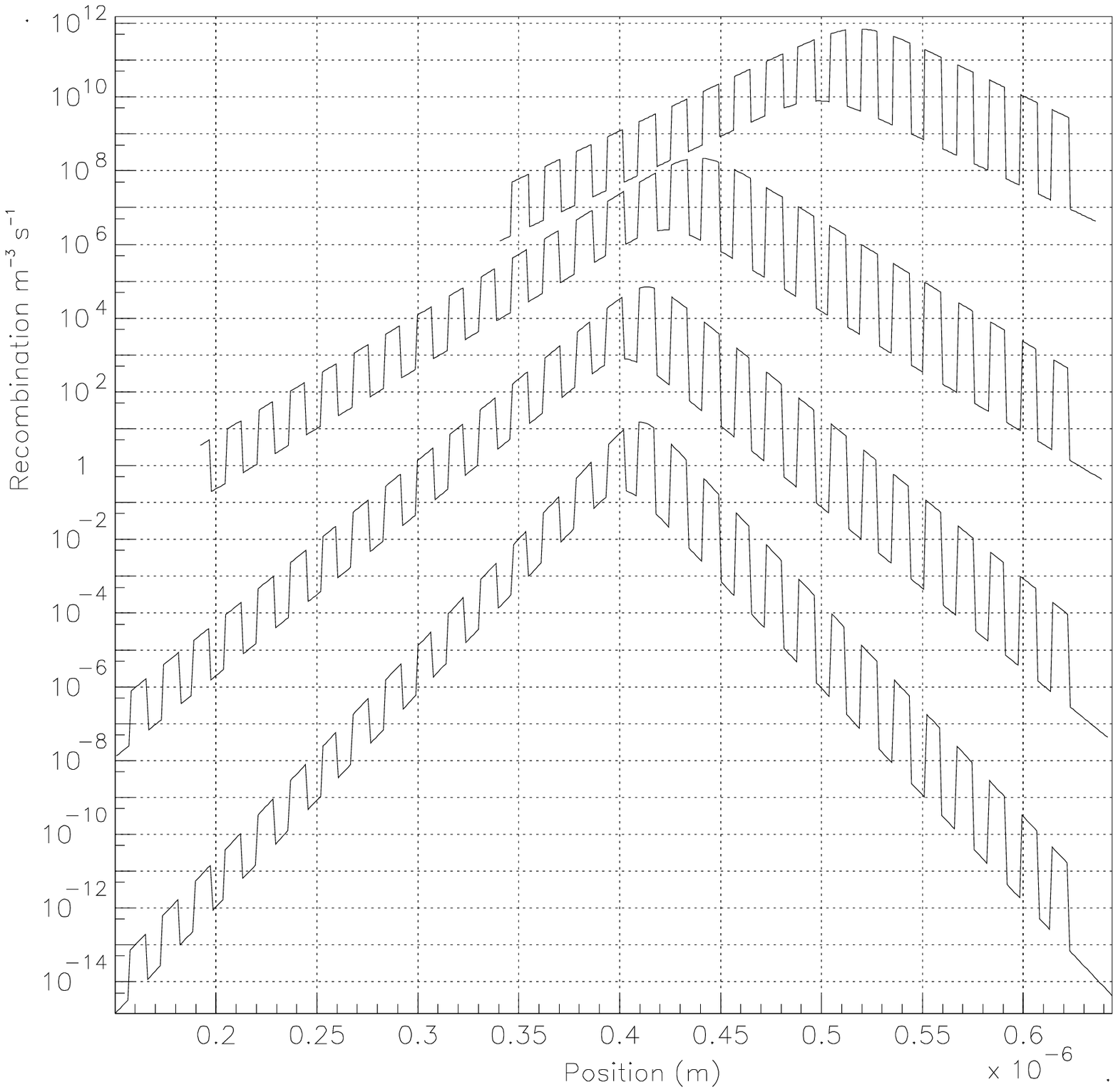}{Shockley-Read-Hall recombination profile 
at low to high bias for a thirty well QWSC showing enhanced 
recombination in the wells and reduced depletion at higher biasses
\label{srhpicture}}

\subsection{Dark current}

The dark current formulation used here expresses the dark current as a 
sum of ideal and SRH contributions.  The ideal diode contribution is 
determined by charge transport in the neutral n and p regions which is 
determined by fitting the spectral response of these layers.  We will 
see that this ideal contribution is negligible except at high forward 
bias in some samples.

As described in more detail in ref.  \cite{connollyalaska}, the SRH 
contribution to the dark current is the integral of the SRH 
recombination rate across the intrinsic region where it is non 
negligible because of significant populations of both electrons and 
holes.  This method is applicable to MQW systems because the large 
number of wells removes the sensitivity of the dark current to well 
position, as opposed to the SQW case.  The SRH rate for quantum wells 
can be expressed \cite{nelson99} in terms of the well understood 
quantum well DOS.

The SRH rate depends on the cell dimensions and material which 
are well known from the epitaxial growth.  Given the knowledge of the 
DOS this leaves the hole and electron lifetimes in the intrinsic 
regions as free parameters in the absence of direct measurement.  If 
electron and hole lifetimes are made significantly different in the 
material being considered (well or barrier), the slope of the dark 
current on a log-linear scale (the ideality) changes significantly in 
contradiction with experiment.

This leaves as two parameters the carrier lifetimes in the barriers, 
and in the wells.  These two values are determined by fitting the dark 
currents of control structures made of material equivalent to wells 
and barriers respectively where they are the sole free parameters.

Given this determination of carrier non radiative lifetimes and the 
knowledge of the DOS, this leaves no free parameters in fitting the 
dark currents of QWSC structures, as will be seen subsequently.

A further assumption we examine is that the Quasi-Fermi level 
separation $\Delta E_{f}$ is constant across the i region and 
determined by the applied bias.  Reducing $\Delta E_{f}$ implies lower 
carrier concentrations and hence lower recombination in the wells as 
investigated below.

Figure \ref{srhpicture} shows a calculated SRH recombination rate
in a 30 well \gaasalgaas\ QWSC with x=36\%. The four curves correspond to
applied biasses ranging from near zero to just below the build-in 
voltage, for the depleted or field bearing section of the intrinsic 
region. The depletion width is reduced as the applied bias is 
increased due to the fixed charge corresponding to unintentional 
background doping.  This background doping is p type in this case, 
which is why the SRH peak shifts towards the n region on the right of 
the graph.

The recombination in the wells is about two orders greater than in the 
barriers in this case and asymetric due to different p and n region 
effective densities of states.  This profile corresponds to a dark 
current which fits the data in figure \ref{ivalgaasmqw} well showing a 
reduced $\Delta E_{f}$ in SRH dominated material as discussed below.

\section{Comparison with data}

\subsection{AlGaAs}

Figure \ref{ivgaascontrols} shows theory and experiment for two GaAs 
p-i-n well controls of different dimensions from different growth runs 
to check repeatability.  The ideal diode Shockley dark current 
component (dashed line) is negligible except at high bias where it 
starts to have an effect.  The dark currents of the two cells are 
essentially identical within errors, but the model requires a slightly 
longer lifetime of 11ns in the wider cell versus 8ns in the narrower.  
Together the results from these different cells show a good 
consistency in model predictions and hence material reproducibility.

\smalleps{htbp}{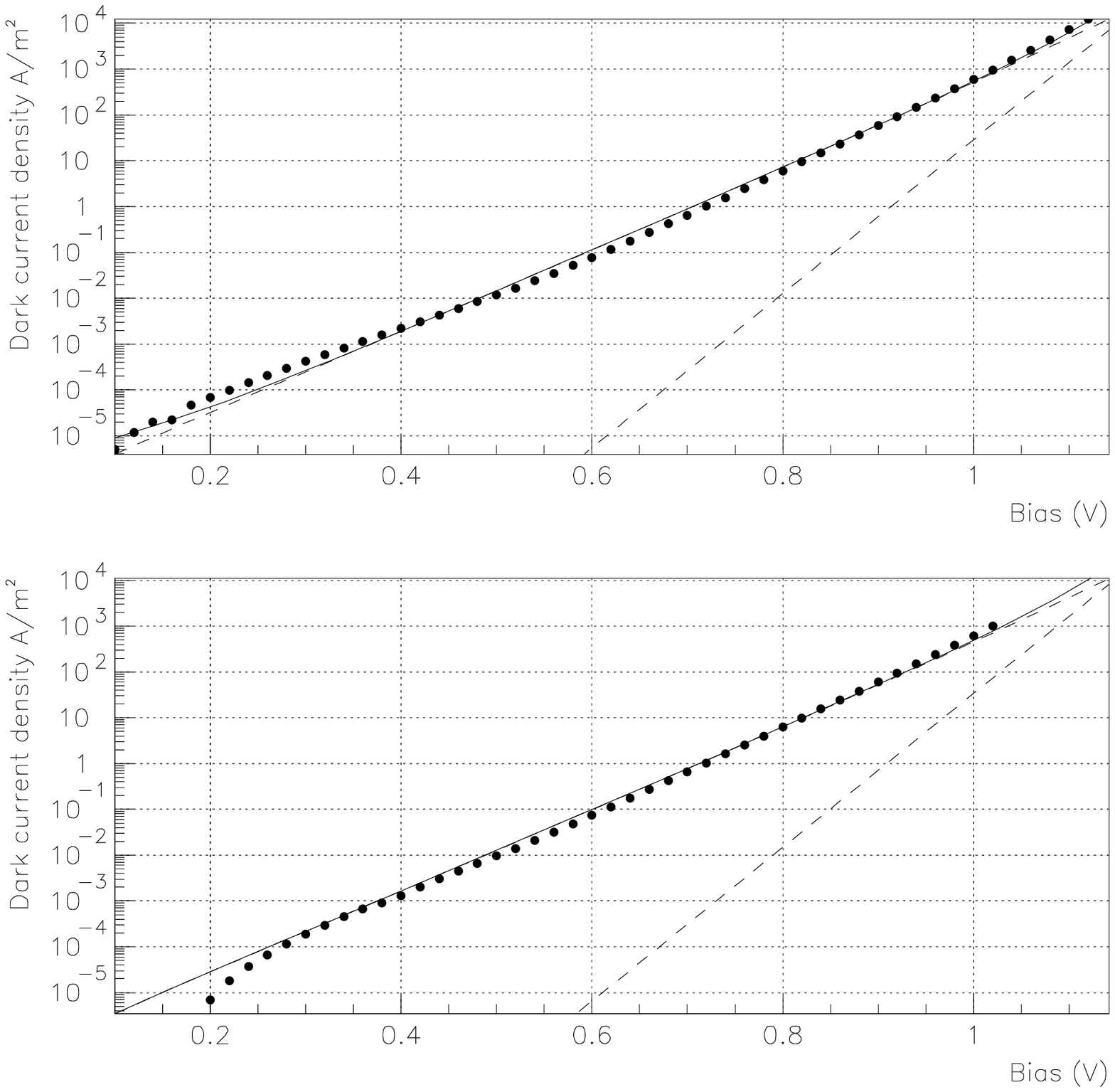}{Theory and experimental dark current for 
a Gaas MQW control with 1$\mu m$ wide i region (upper figure) and 
0.9$\mu m$ wide i region giving non radiative lifetimes of 8ns and 
11ns for undoped GaAs
\label{ivgaascontrols}}

\smalleps{htbp}{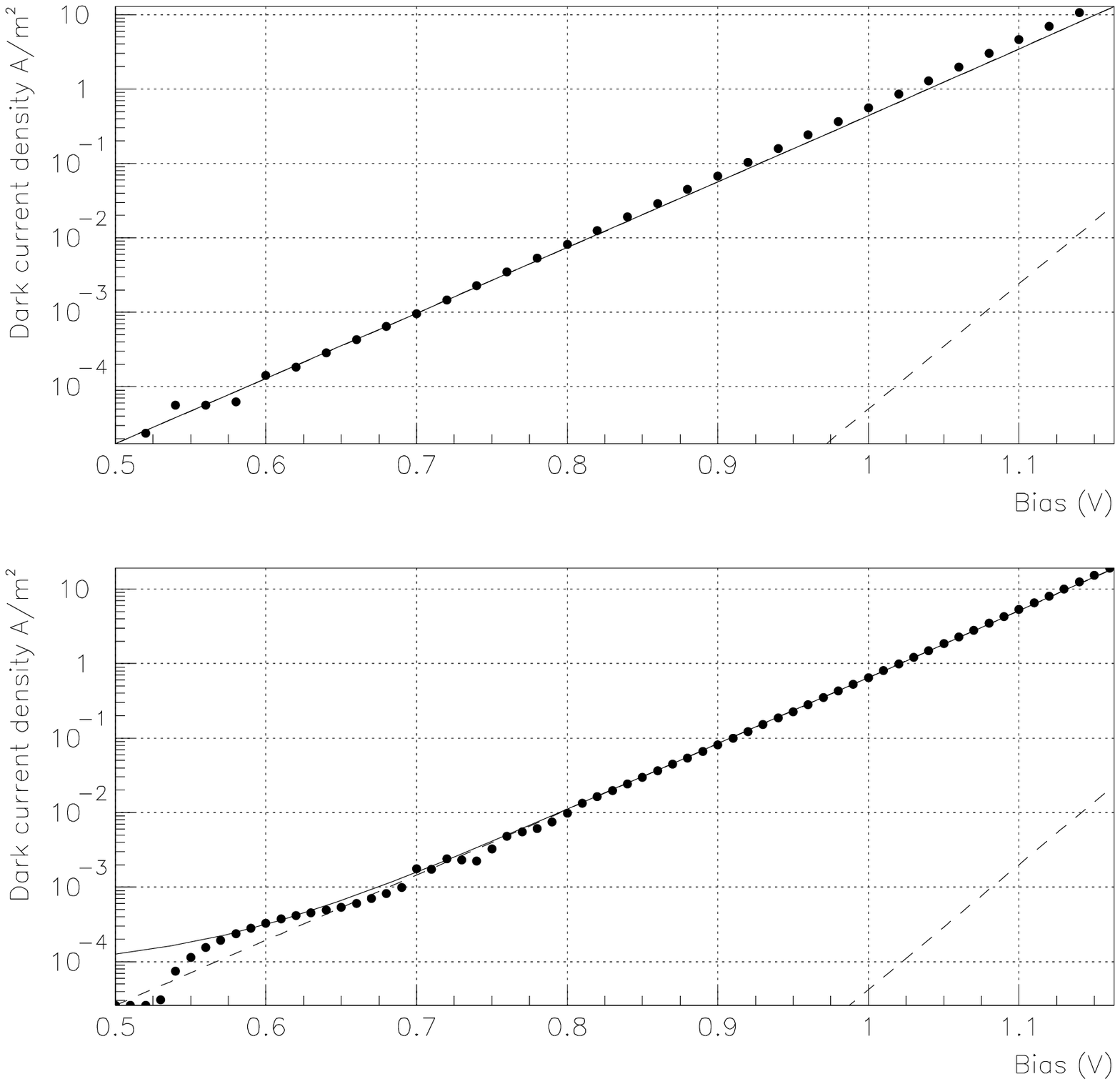}{Theory and experimental dark current 
for an AlGaAs MQW control with 0.48$\mu m$ wide i region (upper 
figure) and SQW 0.31$\mu m$ wide i region control giving non radiative 
lifetimesw of 0.7ns and 0.3ns for undoped $\rm Al_{36}Ga_{0.64}As$
\label{ivalgaascontrols}}

Figure \ref{ivalgaascontrols} shows similar results for two AlGaAs 
controls from different growth runs some time apart with the thicker 
MQW control i region 55\% wider than the narrower SQW control cell.  
The these controls are also consistent, and determine lifetimes of 
0.7ns and 0.3ns in undoped $\rm Al_{36}Ga_{0.64}As$.  Again the ideal 
component is negligible in this case and lower due to different p and 
n layer characteristics.

The consistency of these values gives us confidence in applying these 
lifetimes to material in QWSCs.  We use the MQW control lifetimes (8ns 
for GaAs and 0.7ns for AlGas).  Two examples of modelling \gaasalgaas\ 
QWSC structures are shown in figure \ref{ivalgaasmqw}.  The QWSCs were 
again grown in different growth runs and in different institutions and 
have a significantly different structure.  One features 50 wells and a 
i region $\rm 0.81\mu m$ wide, the other 30 wells in $\rm 0.48\mu m$ i 
region.

\smalleps{htbp}{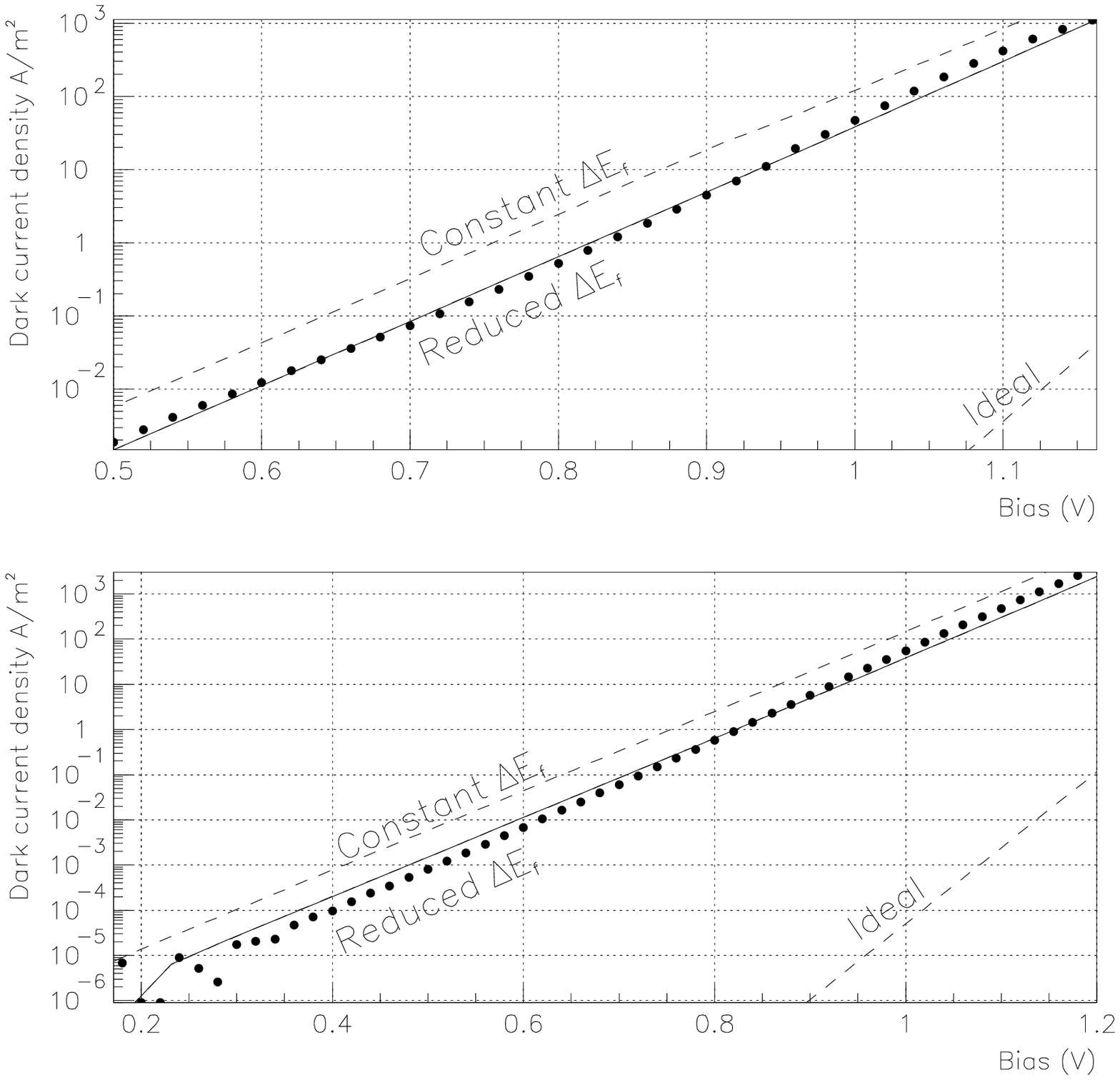}{Theory and experimental dark current for 
two AlGaAs QWSCs with different i region widths and number of wells, 
using the same $\Delta E_{f}$ in the wells \label{ivalgaasmqw}}

The two graphs show the methodology applied to the two samples, with a
fit to the data (dots) assuming a constant $\Delta E_{f}$ (dashed line), and
the same with a reduced $\Delta E_{f}$ which agrees closely with the data.
The value of $\Delta E_{f}$ is 140meV in both cases.

This value is larger than the value reported in ref.  
\cite{nelsonjap97} for SRH dominated SQW samples and developed in ref.  
\cite{nelson99}.  A difference is not surprising given the different 
approaches and different MQW and SQW samples.  Moreover, ref.  
\cite{nelsonjap97} shows that the depletion approximation for SRH 
dominated material is good for biasses up to the operating voltage, 
and is confirmed here since we reach the same conclusion with similar 
values in MQW and SQW samples.  This is that the $\Delta E_{f}$ is 
reduced in QWSC structures, as demonstrated in SQWs.

For this cell the constant $\Delta E_{f}$ calculation predicts an AM1.5 
efficiency 14\% above what might be expected for this cell based on a
constant $\Delta E_{f}$, in agreement with experimental.  The fact 
that this cell does not appear to follow the constant $\Delta E_{f}$ 
picture therefore yields a significant increase in efficiency in this 
system over what might be expected from the more simple prediction.

\subsection{InGaAsP}
Figure \ref{ivingaasp} shows the same analysis for the dark currents 
of QWSCs in \ingaasp\ on InP substrates.  The upper and lower curves 
are well and barrier controls, made of InGaAsP lattice matched to InP 
and are analogues of graphs \ref{ivgaascontrols} and 
\ref{ivalgaascontrols}.

The well control is a double heterostructure control with an i region 
made of material with the well effective bandgap including 
confinement.  The barrier control is a double heterostructure control 
with an i region made of barrier material.  In this case the bulk well 
control lifetime of 70ns is shorter than that of the barrier control 
which is 400ns, but the conclusions are the same.

The middle curves show modelling of the QWSC dark current with the 
lifetimes derived from the two controls.  The dashed line again shows 
the expected dark current with a constant $\Delta E_{f}$ given by the 
applied bias.

The fit requires a reduction in $\Delta E_{f}$ level of 130meV in
the well. The ideal component is again negligible.

This shows similar behaviour to the AlGaAs case with a slightly 
reduced $\Delta E_{f}$ consistent with the slightly shallower wells 
visible in the spectral response fits in figure \ref{qealgaasqwsc} 
and \ref{qeingaaspqwsc}. The sum of well depths in the \gaasalgaas\
case is approximately 400meV versus 290meV in the \ingaasp\ case
including confinement. Furthermore, \ingaasp\ well have
different hole to electron band offsets of 16\% versus about 33\% in
\gaasalgaas\ ith the valence well being deeper.

This shows similar conclusions in a second material despite 
significantly different band structures and carrier properties, and a 
narrower voltage range because of the lower built in voltage. Although 
there is uncertainty in the value of $\Delta E_{f}$ due to the theoretical 
approximations used, the reduction confirms the high efficiency potential of 
QWSCs.

\smalleps{htbp}{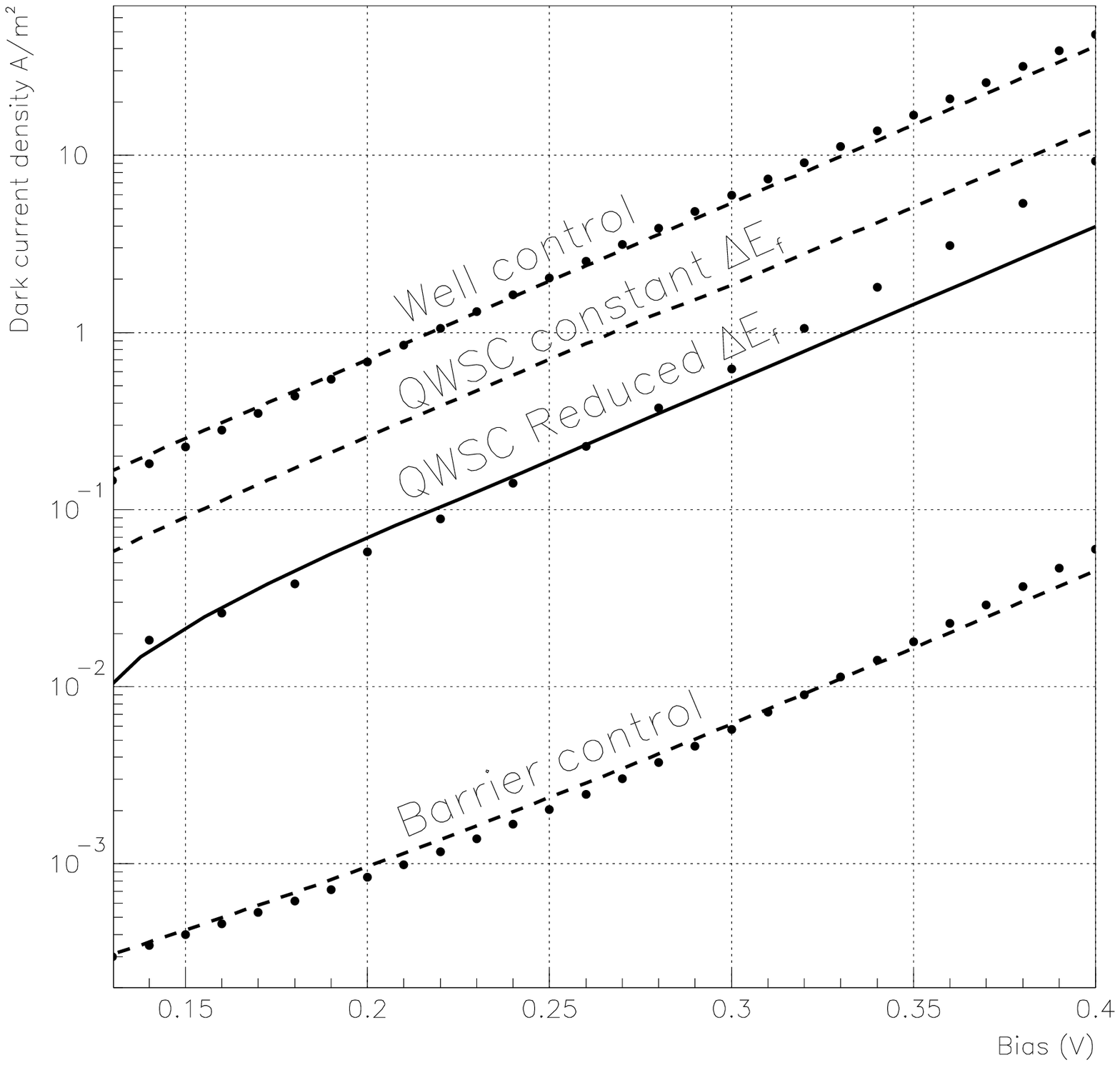}{Theory and experimental dark 
current for a \ingaasp\ MWQ (centre curves) and well and barrier 
controls showing a similar dark current reduction to \gaasalgaas\
QWSCs \label{ivingaasp}}

\section{Conclusions}

The QWSC benefits from an increase in photogeneration in the wells
but suffers from increased recombination in the lower bandgap well
regions. In order to study which effect is greater we study
photocurrent and dark current and express the modifications to dark 
and photocurrent in terms of the quantum well density of states. The 
photocurrent from the wells is determined with no free parameters to 
good accuracy.

The dark current for the control homojunctions is well understood.  
Modelling these gives us an estimate of carrier lifetimes in the 
depletion region, together with verifying the transport parameters 
used in the SR calculation at the onset of ideal diode behaviour, if 
present.

The QWSC dark current depends on four lifetimes. We reduce 
this to two by assuming that hole and electron non radiative lifetimes
are equal. In the absence of direct measurements we derive the barrier 
lifetimes from controls with the barrier composition. Similarly, 
controls with the well composition set an upper limit on the well 
lifetimes employed.

A consistent set of lifetimes in two materials emerges despite differences
in growth and device design which gives confidence in the lifetimes used.

We see a systematic overestimation of the dark current.  This can be 
explained in terms of a smaller value of $\Delta E_{f}$ in the wells.  
This strongly suggests that these structures have the potential to be 
efficient solar cells.

The treatment used here does not apply to SQW structures because of 
the approximations made.  For SQW samples the position of the quantum 
wells and the background doping are significant and require an exact 
solution satisfying Poison's equation.

We note however that the treatment applies reliably to MQW systems 
studied here since dark current is determined mainly by the balance 
between QW and barrier material, in that on average the position of 
the wells is not critical.  Furthermore the depletion approximation is 
reliable up to the operating voltage \cite{nelsonjap97}.  This is 
borne out by the range of control and QWSC \gaasalgaas\ and \ingaasp\ 
samples examined and comparison with previous exact solutions.

Finally we see consistently similar effects in two systems with
very different materials parameters that can explain
the higher efficiency observed in the QWSC system.

\end{document}